\documentclass{elsart}
\usepackage{amssymb}
\usepackage{amsfonts}
\usepackage{amsmath}
\usepackage{graphicx}

\begin{document}

\begin{frontmatter}
\title{On the integrable gravity coupled to fermions}
\author{Vladimir A. Belinski}
\address{ICRANet, 65122 Pescara, Italy and\\IHES, F-91440 Bures-sur-Yvette, France}

\begin{abstract}
In the present letter we indicate an extension of the pure gravity inverse
scattering integration technique to the case when fermions (introduced on
the base of supersymmetry) are present. In this way the integrability
technique for simple ($N=1$) supergravity in two space-time dimensions
coupled to the matter fields taking values in the Lie algebra of $E_{8\left(
+8\right) }$ group is developed. This theory contains matter living only in
one Weyl representation of $SO\left( 16\right) $ and represents the
reduction to two dimensions of the three-dimensional simple supergravity
constructed in \cite{NR}.

Our spectral linear problem use superspace and covers the complete set of
principal bosonic and fermionic equations of motion. This linear system, as
in pure gravity, contains only the first order poles with respect to the
spectral parameter. The procedure of constructing the exact super-solitonic
solutions is outlined.
\end{abstract}

\end{frontmatter}
%\maketitle

\section{ \ \ \ Introduction}

The three-dimensional maximal $(N=16)$ supergravity coupled to the matter
fields taking values in the Lie algebra of $E_{8\left( +8\right) }$ group
has been constructed by N. Marcus and J. Schwarz in \cite{MS}. The theory
has 128 space-time physical scalars parametrizing the coset $E_{8\left(
+8\right) }$ $/$ $SO\left( 16\right) $ and their 128 space-time spinorial
super-partners. From the point of view of the internal space [where $%
SO\left( 16\right) $ symmetry acts] both sets of fields are spinors in such
a way that those which are space-time scalars belong to the "positive"
chirality representation of $SO\left( 16\right) $ and their super-partners
(which are space-time spinors) are in the inequivalent "negative" chirality
representation. This arrangement excludes a possibility to have any
truncation to a theory with all fields in one Weyl representation of $%
SO\left( 16\right) $. However, H. Nishino and S. Rajpoot \cite{NR} have been
pointed out that in three-dimensional space-time such common chirality case
also exists though only for the simple $(N=1)$ supergravity. This
possibility corresponds to quite a new theory which cannot be obtained as
particular case from the Marcus-Schwarz construction.

However, it is easy to check that Lagrangian for the $N=1$ Nishino-Rajpoot
theory (without a cosmological constant) can be obtained from Lagrangian of
the $N=16$ Marcus-Schwarz supergravity by the following formal rules:

a) among $16$ Marcus-Schwarz gravitinos $\psi_{\mu}^{I}$ ($I=1,2,...,16)$
take only one (for example for $I=1$) to be non-zero, \ 

b) among all internal $\Gamma^{I}$-matrices of $SO\left( 16\right) $ (which
are symmetric and off-diagonal) chose the first one (again for $I=1$) to
have the upper-right block as unit matrix,

c) in all quantities (apart of $\Gamma^{I}$-matrices) each dotted index
replace by undotted.

This means that also the reduction of the three-dimensional Nishino-Rajpoot
supergravity to two dimensions can be derived from the reduction of the
Marcus-Schwarz theory to two dimensions using the same three rules.
Fortunately, the reduction of the Marcus-Schwarz theory to two dimensions is
known and this is two-dimensional integrable $N=16$ supergravity constructed
by H. Nicolai \cite{N1}. Hence, the equations of motion of the matter fields
in the dimensionally reduced $N=1$ Nishino-Rajpoot supergravity follow from
the Nicolai $N=16$ two-dimensional supergravity after one apply to the last
theory the operations a), b), c). This formal procedure being applied to the
equations (8) and (11) of the letter \cite{N1} gives\footnote{%
In letter \cite{N1} there is a misprint with the numerical coefficient in
front of the right hand side of equation (8) corresponding to our (\ref{New1}%
). The correct value for this coefficient (namely $i/8$) have been found by
Nicolai and Warner later in \cite{NW}. The numerical coefficients in front
of the right hand side of our equation (\ref{New2}) and in corresponding
equation (11) in letter \cite{N1} both are correct. The difference ($1/16$
and $1/24$) is due to the fact that in the r.h.s. of our (\ref{New2}) we
used the sum $\gamma_{\alpha}...\gamma^{\alpha}...$ only over $\alpha=0,1$
while in \cite{N1} this summation was extended up to the addend $%
\gamma_{3}...\gamma^{3}...$, where $\gamma_{3}=\gamma^{3}=\gamma_{0}%
\gamma_{1}$. However, the difference manifests itself only in the
aforementioned overall factors.
\par
In our letter we don't use the special designation $D_{\alpha}$ for
covariant (with respect to the local orthogonal rotations of the frame $%
\mathbf{V}$\textbf{) }derivatives. In all our equations such derivatives are
written in the explicit forms.} the following equations of motion for the
matter fields (that is for $128$ scalars $\varphi^{A}$ and $128$ spinors $%
\chi^{A}$) in the two-dimensional version of the Nishino-Rajpoot theory: 
\begin{equation}
\frac{1}{\alpha}\eta^{\alpha\beta}\left[ (\alpha p_{\alpha}^{A})_{,\beta }+%
\frac{1}{4}q_{\alpha}^{IJ}\Gamma_{AB}^{IJ}\alpha p_{\beta}^{B}\right] =\frac{%
i}{8}\Gamma_{AB}^{IJ}p_{\alpha}^{B}\bar{\chi}^{C}\gamma^{\alpha}%
\Gamma_{CD}^{IJ}\chi^{D}\text{ },  \label{New1}
\end{equation}%
\begin{equation}
-i\frac{1}{\sqrt{\alpha}}\gamma^{\alpha}\left[ (\sqrt{\alpha}\chi
^{A})_{,\alpha}+\frac{1}{4}q_{\alpha}^{IJ}\Gamma_{AB}^{IJ}\gamma^{\alpha}%
\sqrt{\alpha}\chi^{B}\right] =\frac{1}{16}\gamma^{\alpha}\Gamma_{AB}^{IJ}%
\chi^{B}\bar{\chi}^{C}\gamma_{\alpha}\Gamma_{CD}^{IJ}\chi^{D}\text{ }.
\label{New2}
\end{equation}
All notations used here are explained in the section 3 of our letter (they
are the same as in papers \cite{MS, N1, NR}).

Remarkably enough the equations (\ref{New1})-(\ref{New2}) obtained by the
above-mentioned formal rules coincide exactly with those derived in exact
mathematical way from our superspace linear spectral problem as its
self-consistency conditions (see section 3). Consequently, these equations
are integrable and exact solitonic type solutions for them can be
constructed.

A question might arises why one need to go to superspace to search a new
linear spectral problem and cannot extract it from that one already
presented in \cite{N1} with the aid of the same rules a), b), c)? Such a
way, however, will give a Lax pair which will not be able to cover the
fermionic equations (\ref{New2}). The reason is that the linear system
proposed in \cite{N1} is not complete in the sense that fermionic equation
(11) of the letter \cite{N1} [that one which generates our (\ref{New2})] do
not follow from this linear system and must be added by hands. This
circumstance is evident from the fact that the Lax pair of \cite{N1}
contains only quadratic products of anticommuting fermions and corresponding
self-consistency conditions cannot produce any equation of motion containing
terms of the odd powers of such fermions. [Besides there is one undesirable
point (although not of any principal significance) in relation to the Lax
pair constructed in \cite{N1} which lead to some technical complications.
This is appearance the poles of the second order with respect to the
spectral parameter while the corresponding spectral problem in pure gravity
has only simple poles]. The approach proposed in the present letter dismiss
these nuisances.

\section{ \ The multidimensional version of the integrable gravity}

It is known \cite{BZ, BV} that in any space-time of dimension $n+2$ with
interval 
\begin{equation}
ds^{2}=g_{\alpha \beta }\left( x^{0},x^{1}\right) dx^{\alpha }dx^{\beta
}+g_{ik}\left( x^{0},x^{1}\right) dy^{i}dy^{k}  \label{M1}
\end{equation}%
(where $\alpha ,\beta ,\gamma ,...=0,1$ and $i,k,l,...=1,2,...n$) the
Einstein equations are integrable since the equations for the metric
coefficients $g_{ik}\left( x^{0},x^{1}\right) $ follows from the Lax pair as
its self-consistency conditions and the metric components $g_{\alpha \beta
}\left( x^{0},x^{1}\right) $ can be found by quadratures in terms of the
known $g_{ik}\left( x^{0},x^{1}\right) .$ Without loss of generality the
2-dimensional block $g_{\alpha \beta }\left( x^{0},x^{1}\right) $ can be
chosen in conformal flat form:%
\begin{equation}
g_{\alpha \beta }=\lambda ^{2}\eta _{\alpha \beta }\text{ },\text{ }\eta
_{\alpha \beta }=diag\left( \eta _{00},\eta _{11}\right) =diag\left(
1,-1\right) \text{ }.  \label{M2}
\end{equation}%
\ It is convenient to introduce matrix $\mathbf{G}$ (with entries $G_{ik}$)%
\footnote{%
All multidimensional matrices (apart from the two-dimensional Dirac
gamma-matrices) we designate by the bold letters. Tilde at the top of a
matrix means transposition. The functions which depend also on the spectral
parameter we designate by the letters with the hat on the top. The simple
partial derivatives are denoted by comma.} and represent the metric matrix $%
\mathbf{g}$ (with entries $g_{ik})$ as%
\begin{equation}
\mathbf{g}=\alpha ^{2/n}\mathbf{G}\text{ },\text{ }\det \mathbf{G}=1\text{ },%
\text{ }\det \mathbf{g}=\alpha ^{2}.  \label{M3}
\end{equation}%
Then the components $R_{\alpha i}$ of the Ricci tensor vanish identically
and equations $R_{\alpha \beta }=0$ and $R_{ik}=0$ can be written in matrix
form using the light-like variables $\zeta ,\eta $:%
\begin{equation}
x^{0}=\eta +\zeta \text{ },\text{ }x^{1}=\eta -\zeta \text{ }.  \label{M4}
\end{equation}%
Equations $R_{ik}=0$ are:%
\begin{equation}
\alpha _{,\zeta \eta }=0\text{ },  \label{M5}
\end{equation}%
\begin{equation}
(\alpha \mathbf{G}^{-1}\mathbf{G}_{,\zeta })_{,\eta }+(\alpha \mathbf{G}^{-1}%
\mathbf{G}_{,\eta })_{,\zeta }=\mathbf{0}.  \label{M6}
\end{equation}%
Equations $R_{\alpha \beta }=0$ are equivalent to the system\footnote{%
In fact the system $R_{\alpha \beta }=0$ contains one more equation
additional to the (\ref{M7})-(\ref{M8}) which is of second order for $f$ \
but it is the direct consequence of (\ref{M5})-(\ref{M8}) and we can forget
about it.}:%
\begin{equation}
\frac{f_{,\zeta }}{f}=\frac{\alpha }{4\alpha _{,\zeta }}Tr\left[ \left( 
\mathbf{G}^{-1}\mathbf{G}_{,\zeta }\right) ^{2}\right] \text{ },  \label{M7}
\end{equation}%
\begin{equation}
\frac{f_{,\eta }}{f}=\frac{\alpha }{4\alpha _{,\eta }}Tr\left[ \left( 
\mathbf{G}^{-1}\mathbf{G}_{,\eta }\right) ^{2}\right] \text{ },  \label{M8}
\end{equation}%
where%
\begin{equation}
f=\frac{\alpha ^{\left( n-1\right) /n}\lambda ^{2}}{\alpha _{,\zeta }\alpha
_{,\eta }}\text{ }.  \label{M9}
\end{equation}%
If equations (\ref{M5}) and (\ref{M6}) are solved\footnote{%
The following important point should be stressed here. From the relations (%
\ref{M3}) follows that solution of equation (\ref{M6}) for the matrix $%
\mathbf{G}$ should satisfy the restriction $\det \mathbf{G=1.}$ However, the
application of the inverse scattering integration procedure to the equation (%
\ref{M6}), if taking this restriction from the outset, technically are a
little bit awkward. More convenient approach is to solve (\ref{M6}) [under
given solution of (\ref{M5}) for $\alpha $] first ignoring any additional
restricton for the $\det \mathbf{G}$ but at the end of calculation make the
simple rescaling of the solution in order to get the necessary condition of
the unit determinant\textbf{. }The trick is as follows. If we obtained the
solution of the equation (\ref{M6}) with $\det \mathbf{G\neq 1}$ we can pass
to the new "physical" matrix $\mathbf{G}_{\left( ph\right) }$ by the
transformation $\mathbf{G}_{\left( ph\right) }=\left( \det \mathbf{G}\right)
^{-1/n}\mathbf{G}$. It is simple task to prove that the matrix $\mathbf{G}%
_{\left( ph\right) }$ is also a solution of the same equation (\ref{M6}) and
simultaneously satisfy the condition $\det \mathbf{G}_{\left( ph\right) }=1$.%
} then $f$ can be found by quadratures from equations (\ref{M7}) and (\ref%
{M8}) [the self-consistency conditions of which are satisfied automatically
if $\alpha $ and $\mathbf{G}$ are the solutions of the equations (\ref{M5})
and (\ref{M6})].

The spectral linear problem associated with the main equation (\ref{M6}) we
used in \cite{BZ} contains the differentiation also with respect to the
spectral parameter but this is an inessential technical point. Our original
Lax representation for the equation (\ref{M6}) can be written also in the
following equivalent form:%
\begin{equation}
\mathbf{\hat{G}}^{-1}\mathbf{\hat{G}}_{,\zeta}=\frac{\alpha}{\alpha -s}%
\mathbf{G}^{-1}\mathbf{G}_{,\zeta}\text{ },\text{ }\mathbf{\hat{G}}^{-1}%
\mathbf{\hat{G}}_{,\eta}=\frac{\alpha}{\alpha+s}\mathbf{G}^{-1}\mathbf{G}%
_{,\eta}\text{ },\text{ }  \label{M10}
\end{equation}
where $\mathbf{\hat{G}}\left( \zeta,\eta,s\right) $ depends on the complex
spectral parameter $s$ which depends on the coordinates $\zeta,\eta$ in
accordance with differential equations:%
\begin{equation}
\frac{s_{,\zeta}}{s}=\frac{2\alpha_{,\zeta}}{\alpha-s}\text{ },\text{ }\frac{%
s_{,\eta}}{s}=\frac{2\alpha_{,\eta}}{\alpha+s}\text{ }.  \label{M11}
\end{equation}
The self-consistency requirement for the last equations is satisfied due to
the condition (\ref{M5}). The solution of the equations (\ref{M11}) contains
one arbitrary complex constant $w$ then parameter $s=s(\zeta,\eta,w)$ has
one arbitrary degree of freedom independent of those due to the changing of
coordinates. This means that in the integrability conditions for the pair (%
\ref{M10}) all terms containing the different powers of $s$ must vanish
separately. The matrix $\mathbf{G}$ in the right hand side of (\ref{M10}) is
a function on the two coordinates $\zeta$ and $\eta$ only, that is it is
treated as unknown "potential" independent on the parameter $s$. The
function $\alpha(\zeta,\eta)$ which is a solution of the wave equation (\ref%
{M5}) should be considered in (\ref{M10}) as some given external field. The
equation of interest (\ref{M6}) should result from the linear spectral
system (\ref{M10}) as its self-consistency (integrability) conditions and it
is easy to check that this is indeed the case.

For any regular at the point $s=0$ solution $\mathbf{\hat{G}}\left(
\zeta,\eta,s\right) $ of the Lax pair (\ref{M10}) the solution of the
equation (\ref{M6}) for matrix $\mathbf{G}\left( \zeta,\eta\right) $ follows
automatically from the relation:%
\begin{equation}
\mathbf{G}\left( \zeta,\eta\right) =[\mathbf{\hat{G}}\left( \zeta
,\eta,s\right) ]_{s=0}\text{ }.  \label{M11-1}
\end{equation}
In case of solitonic fields the procedure of integration of the spectral
linear problem (\ref{M10}) for matrix $\mathbf{\hat{G}}\left( \zeta
,\eta,s\right) $ consists of the following steps. First we need to have some
background solutions $\alpha(\zeta,\eta)$ and $\mathbf{G}_{0}(\zeta,\eta)$
of the gravitational equations (\ref{M5})-(\ref{M6}) and then to find from
equations (\ref{M10}) the corresponding background spectral matrix $\mathbf{%
\hat{G}}_{0}\left( \zeta,\eta,s\right) .$ After that we "dress" $\mathbf{%
\hat{G}}_{0}\left( \zeta,\eta,s\right) $ by the simplest meromorphic (having
only isolated first-order poles with respect to the spectral parameter $s$)
matrix $\mathbf{\hat{K}}\left( \zeta,\eta,s\right) $, that is we represent $%
\mathbf{\hat{G}}$ in the form $\mathbf{\hat{G}}\left( \zeta,\eta,s\right) =$ 
$\mathbf{\hat{G}}_{0}\left( \zeta ,\eta,s\right) \mathbf{\hat{K}}\left(
\zeta,\eta,s\right) $. Substituting this form into equations (\ref{M10}) we
can find the matrix $\mathbf{\hat{K}}\left( \zeta,\eta,s\right) $ in terms
of the known background solution $\mathbf{\hat{G}}_{0}\left(
\zeta,\eta,s\right) $ and given wave function $\alpha(\zeta,\eta)$ by pure
algebraic procedure (namely this is the principal advantage of the method).
The number of poles with respect to the spectral parameter $s$ in matrix $%
\mathbf{\hat{K}}\left( \zeta,\eta,s\right) $ is the number of solitons we
add to the background. Then the final solution of interest for $\mathbf{G}%
(\zeta,\eta)$ follows from the relation (\ref{M11-1}).

\section{Generalization to the case when fermions are present}

Generalization of the scheme described above to the case when a number of
fermionic fields are present can be done by introducing the superspace with
coordinates $x^{0},x^{1},\theta^{1},\theta^{2}$ where $\theta^{1}$ and $%
\theta^{2}$ are odd (anticommuting) variables. Now the spectral matrix $%
\mathbf{\hat{G}}\left( \zeta,\eta,s\right) $ which appeared in the equations
(\ref{M10}) should be replaced by a supermatrix $\mathbf{\hat{\Psi}}\left(
\zeta,\eta,\theta^{1},\theta^{2},s\right) $ with even (commuting) entries
and instead of the simple derivatives $\partial_{\zeta}$ and $%
\partial_{\eta} $ in (\ref{M10}) we should use the following odd
differential operators:%
\begin{equation}
D_{\zeta}=\frac{\partial}{\partial\theta^{2}}-\theta^{2}\frac{\partial }{%
\partial\zeta}\text{ },  \label{M12-1}
\end{equation}%
\begin{equation}
D_{\eta}=-\frac{\partial}{\partial\theta^{1}}+\theta^{1}\frac{\partial }{%
\partial\eta}\text{ }.  \label{M12-2}
\end{equation}
These operators anticommute:%
\begin{equation}
D_{\zeta}D_{\eta}+D_{\eta}D_{\zeta}=0\text{ },  \label{M14}
\end{equation}
and the superspace generalization of the Lax pair (\ref{M10}) becomes:%
\begin{equation}
\mathbf{\hat{\Psi}}^{-1}D_{\zeta}\mathbf{\hat{\Psi}}=\frac{\alpha}{\alpha -s}%
\mathbf{\Psi}^{-1}D_{\zeta}\mathbf{\Psi}\text{ },\text{ }\mathbf{\hat{\Psi }}%
^{-1}D_{\eta}\mathbf{\hat{\Psi}}=\frac{\alpha}{\alpha+s}\mathbf{\Psi}%
^{-1}D_{\eta}\mathbf{\Psi}\text{ },  \label{M15}
\end{equation}
where $\alpha(x^{0},x^{1})$ still is an usual even function which does not
depends on the $\theta$-coordinates and satisfies the wave equation (\ref{M5}%
). The spectral parameter $s(x^{0},x^{1})$ also is even and does not depends
on the $\theta$-coordinates and follows from the differential equations (\ref%
{M11}) after function $\alpha(x^{0},x^{1})$ is fixed.

For any regular at the point $s=0$ solution of equations (\ref{M15}) for $%
\mathbf{\hat{\Psi}}$ the matrix of interest $\mathbf{\Psi }$ comes from the
relation:%
\begin{equation}
\mathbf{\Psi }\left( \zeta ,\eta ,\theta ^{1},\theta ^{2}\right) \mathbf{=[%
\hat{\Psi}}\left( \zeta ,\eta ,\theta ^{1},\theta ^{2},s\right) ]_{s=0}\text{
}.  \label{M16}
\end{equation}%
The crucial point is that the self-consistency condition $D_{\zeta }\left(
D_{\eta }\mathbf{\hat{\Psi}}\right) +D_{\eta }\left( D_{\zeta }\mathbf{\hat{%
\Psi}}\right) =\mathbf{0}$ of the generalized Lax pair (\ref{M15}) reduces
only to one supermatrix equation [superspace analog of (\ref{M6})]: 
\begin{equation}
D_{\zeta }(\alpha \mathbf{\Psi }^{-1}D_{\eta }\mathbf{\Psi )-}D_{\eta
}(\alpha \mathbf{\Psi }^{-1}D_{\zeta }\mathbf{\Psi )=0}\text{ },  \label{M17}
\end{equation}%
which is equivalent to the set of equations of motion for the bosonic and
fermionic component fields which are represented by the coefficients in the
expansion of $\mathbf{\Psi }\left( \zeta ,\eta ,\theta ^{1},\theta
^{2}\right) $ with respect to the $\theta $-coordinates. The matrix $\mathbf{%
\Psi }\left( \zeta ,\eta ,\theta ^{1},\theta ^{2}\right) $ has structure:

\begin{equation}
\mathbf{\Psi}=\mathbf{J(I+}\text{ }\theta^{1}\mathbf{\Omega}_{1}+\theta ^{2}%
\mathbf{\Omega}_{2}+\theta^{1}\theta^{2}\mathbf{H)}\text{ },  \label{M19}
\end{equation}%
\begin{equation}
\mathbf{\Psi}^{-1}=\left[ \mathbf{I-}\theta^{1}\mathbf{\Omega}_{1}-\theta
^{2}\mathbf{\Omega}_{2}-\theta^{1}\theta^{2}\mathbf{H}-\theta^{1}\theta
^{2}\left( \mathbf{\Omega}_{1}\mathbf{\Omega}_{2}-\mathbf{\Omega}_{2}\mathbf{%
\Omega}_{1}\right) \right] \mathbf{J}^{-1},  \label{M19-1}
\end{equation}
where $\mathbf{I}$ is the unity, $\mathbf{J}\left( \zeta,\eta\right) $ and $%
\mathbf{H}\left( \zeta,\eta\right) $ have even entries and both $\mathbf{%
\Omega}_{1}\left( \zeta,\eta\right) $ and $\mathbf{\Omega}_{2}\left(
\zeta,\eta\right) $ consist of the odd entries. If we are able to find the
solution of equations (\ref{M15}) for the spectral matrix $\mathbf{\hat{\Psi}%
}\left( \zeta,\eta,\theta^{1},\theta^{2},s\right) $ (this is the main
problem of the method, however, solvable for the solitonic fields) then the
solution of the equation (\ref{M17}) for matrix $\mathbf{\Psi}\left(
\zeta,\eta,\theta^{1},\theta^{2}\right) $ follows automatically from the
relation (\ref{M16}) together with component fields of interest $\mathbf{J},%
\mathbf{\Omega}_{1},\mathbf{\Omega}_{2},\mathbf{H}$. This means that these
fields will satisfy automatically the differential equations following from (%
\ref{M17}) after we substitute into it expansions (\ref{M19})-(\ref{M19-1}).
These differential equations\footnote{%
Direct derivation of the first integrability condition of the Lax pair (\ref%
{M15}) gives it as $\left( \alpha\mathbf{J}^{-1}\mathbf{J}_{,\zeta}+\alpha%
\mathbf{\Omega}_{2}^{2}\right) _{,\eta}+\left( \alpha\mathbf{J}^{-1}\mathbf{J%
}_{,\eta}+\alpha\mathbf{\Omega }_{1}^{2}\right) _{,\zeta}=0$ $,$ however
using (\ref{M22}) and (\ref{M23}) it can be transformed to the form (\ref%
{M21}).} are:

\begin{gather}
\frac{1}{\alpha}\left( \alpha\mathbf{J}^{-1}\mathbf{J}_{,\zeta}\right)
_{,\eta}+\frac{1}{\alpha}\left( \alpha\mathbf{J}^{-1}\mathbf{J}_{,\eta
}\right) _{,\zeta}=\frac{1}{2}(\mathbf{J}^{-1}\mathbf{J}_{,\zeta }\mathbf{%
\Omega}_{1}^{2}-\mathbf{\Omega}_{1}^{2}\mathbf{J}^{-1}\mathbf{J}_{,\zeta})%
\text{ }  \label{M21} \\
+\frac{1}{2}(\mathbf{J}^{-1}\mathbf{J}_{,\eta}\mathbf{\Omega}_{2}^{2}-%
\mathbf{\Omega}_{2}^{2}\mathbf{J}^{-1}\mathbf{J}_{,\eta})\text{ },  \notag
\end{gather}

\begin{equation}
\frac{1}{\sqrt{\alpha}}(\sqrt{\alpha}\mathbf{\Omega}_{1})_{,\zeta}+\frac{1}{2%
}(\mathbf{J}^{-1}\mathbf{J}_{,\zeta}\mathbf{\Omega}_{1}-\mathbf{\Omega}_{1}%
\mathbf{J}^{-1}\mathbf{J}_{,\zeta})+\frac{1}{4}\left( \mathbf{\Omega}_{2}^{2}%
\mathbf{\Omega}_{1}-\mathbf{\Omega}_{1}\mathbf{\Omega}_{2}^{2}\right) =0%
\text{ },  \label{M22}
\end{equation}%
\begin{equation}
\frac{1}{\sqrt{\alpha}}(\sqrt{\alpha}\mathbf{\Omega}_{2})_{,\eta}+\frac{1}{2}%
(\mathbf{J}^{-1}\mathbf{J}_{,\eta}\mathbf{\Omega}_{2}-\mathbf{\Omega}_{2}%
\mathbf{J}^{-1}\mathbf{J}_{,\eta})+\frac{1}{4}\left( \mathbf{\Omega}_{1}^{2}%
\mathbf{\Omega}_{2}-\mathbf{\Omega}_{2}\mathbf{\Omega}_{1}^{2}\right) =0%
\text{ },  \label{M23}
\end{equation}%
\begin{equation}
\mathbf{H=}\frac{1}{2}\left( \mathbf{\Omega}_{2}\mathbf{\Omega}_{1}-\mathbf{%
\Omega}_{1}\mathbf{\Omega}_{2}\right) \text{ }.  \label{M24}
\end{equation}

As follows from the foregoing the equations (\ref{M21})-(\ref{M23}) are
integrable because they are the self-consistency conditions of the super Lax
pair (\ref{M15}) but what is much more interesting they coincide exactly
with equations of motion (\ref{New1})-(\ref{New2}) for the matter in the
Nishino-Rajpoot supergravity if the fields $\mathbf{J},\mathbf{\Omega }_{1},%
\mathbf{\Omega }_{2}$ take values in the $E_{8\left( +8\right) }$ group. To
show this coincidence explicitly it is necessary to return in the system (%
\ref{M21})-(\ref{M23}) to the Cartesian coordinates $x^{0},x^{1}$ defined in
(\ref{M4}) and to use instead of the metric $\mathbf{J}$ the orthonormal
frame $\mathbf{V}$: 
\begin{equation}
\mathbf{J=V\tilde{V}}.\text{ }  \label{N1}
\end{equation}%
Instead of the matrices $\mathbf{\Omega }_{1},\mathbf{\Omega }_{2}$ it is
convenient to use their similar images $\mathbf{\Lambda }_{1}\mathbf{,%
\mathbf{\Lambda }_{2}\ }$with respect to the frame $\mathbf{V}$:%
\begin{equation}
\mathbf{\Omega }_{1}=\mathbf{\tilde{V}}^{-1}\mathbf{\Lambda }_{1}\mathbf{%
\tilde{V}}\text{ },\text{ }\mathbf{\Omega }_{2}=\mathbf{\tilde{V}}^{-1}%
\mathbf{\Lambda }_{2}\mathbf{\tilde{V}}\text{ }.  \label{N2}
\end{equation}%
The frame current $\mathbf{V}^{-1}\mathbf{V}_{,\alpha }$ can be decomposed
into antisymmetric and symmetric parts:%
\begin{equation}
\mathbf{V}^{-1}\mathbf{V}_{,\alpha }=\mathbf{Q}_{\alpha }+\mathbf{P}_{\alpha 
\text{ \ }},  \label{N3}
\end{equation}%
where matrices $\mathbf{Q}_{\alpha \text{ \ }}$are antisymmetric and
matrices $\mathbf{P}_{\alpha \text{\ }}$symmetric. From (\ref{N1}) and (\ref%
{N3}) we obtain the following useful expression for the metric current $%
\mathbf{J}^{-1}\mathbf{J}_{,\alpha }$:%
\begin{equation}
\mathbf{J}^{-1}\mathbf{J}_{,\alpha }=2\mathbf{\tilde{V}}^{-1}\mathbf{P}%
_{\alpha \text{ \ }}\mathbf{\tilde{V}}\text{ }.  \label{N4}
\end{equation}

Now it is easy to see that equations (\ref{M22}) and (\ref{M23}) have a
natural Dirac spinorial structure in two-dimensional Minkowski space-time
with metric $\eta_{\alpha\beta}$ defined by (\ref{M2}) and with the
following gamma-matrices:%
\begin{equation}
\gamma^{0}=\left( 
\begin{array}{cc}
0 & -i \\ 
i & 0%
\end{array}
\right) \text{ },\text{ }\gamma^{1}=\left( 
\begin{array}{cc}
i & 0 \\ 
0 & -i%
\end{array}
\right) \text{ },\text{ }\gamma_{0}=\gamma^{0}\text{ },\text{ }\gamma
_{1}=-\gamma^{1}.  \label{N5}
\end{equation}
If we introduce the matrix-generalized two-component Dirac-like spinor $%
\mathbf{\Phi}$ 
\begin{equation}
\mathbf{\Phi=}\left( 
\begin{array}{c}
\mathbf{\Phi}_{1} \\ 
\mathbf{\Phi}_{2}%
\end{array}
\right) =\left( 
\begin{array}{c}
\mathbf{\Lambda}_{1}+\mathbf{\Lambda}_{2} \\ 
\mathbf{\Lambda}_{1}-\mathbf{\Lambda}_{2}%
\end{array}
\right) ,  \label{N6}
\end{equation}
in which the components $\mathbf{\Phi}_{1}$ and $\mathbf{\Phi}_{2}$ are
internal matrices of arbitrary size (their internal \ matrix structure has
no relation to the two-dimensional Dirac algebra in two-dimensional
space-time), then in\textbf{\ }terms of such spinor and frame current
components $\mathbf{Q}_{\alpha},\mathbf{P}_{\alpha\text{ \ }}$equations (\ref%
{M21})-(\ref{M23}) take the form:%
\begin{equation}
\eta^{\alpha\beta}\left[ \frac{1}{\alpha}(\alpha\mathbf{P}_{\alpha})_{,\beta
}+\mathbf{Q}_{\alpha}\mathbf{P}_{\beta}-\mathbf{P}_{\beta}\mathbf{Q}_{\alpha
}\right] =\frac{1}{8}\mathbf{P}_{\alpha}\left( \mathbf{\bar{\Phi}}%
\gamma^{\alpha}\mathbf{\Phi}\right) -\frac{1}{8}\left( \mathbf{\bar{\Phi}}%
\gamma^{\alpha}\mathbf{\Phi}\right) \mathbf{P}_{\alpha}\text{ },  \label{N7}
\end{equation}

\begin{gather}
\frac{1}{\sqrt{\alpha}}\gamma^{\alpha}(\sqrt{\alpha}\mathbf{\Phi)}_{,\alpha
}+\mathbf{Q}_{\alpha}(\gamma^{\alpha}\mathbf{\Phi)}-(\gamma^{\alpha }\mathbf{%
\Phi)Q}_{\alpha}=\frac{1}{16}\left( \gamma_{\alpha}\mathbf{\Phi }\right)
\left( \mathbf{\bar{\Phi}}\gamma^{\alpha}\mathbf{\Phi}\right)  \label{N8} \\
-\frac{1}{16}\left( \mathbf{\bar{\Phi}}\gamma^{\alpha}\mathbf{\Phi}\right)
\left( \gamma_{\alpha}\mathbf{\Phi}\right) \text{ },  \notag
\end{gather}
where%
\begin{equation}
\mathbf{\bar{\Phi}=}\left( \mathbf{\Phi}_{1}\text{ },\mathbf{\Phi}%
_{2}\right) \gamma^{0}=i\left( \mathbf{\Phi}_{2}\text{ },-\mathbf{\Phi}%
_{1}\right) .  \label{N9}
\end{equation}

Now we can apply these form of the original system (\ref{M21})-(\ref{M23})
to the case when the fields $\mathbf{J},\mathbf{\Omega}_{1},\mathbf{\Omega}%
_{2}$ parametrize the $E_{8\left( +8\right) }$ group. This means that the
fields $\mathbf{Q}_{\alpha},\mathbf{P}_{\alpha},\mathbf{\Phi}$ are in the
Lie algebra of $E_{8\left( +8\right) }$ that is they can be represented as
superpositions (with local coefficients) of the generators of this algebra
which generators consist of $120$ antisymmetric matrices $\mathbf{X}^{IJ}$ ($%
I,J,K,L=1,2,...,16$) and $128$ symmetric matrices $\mathbf{Y}^{A}$ ($%
A,B,C,D=1,2,...,128$) each of the size $248\times248$ (the concrete
realization of these matrices see, for example, in \cite{CPS}). The defining
relations are:%
\begin{equation}
\left[ \mathbf{X}^{IJ},\mathbf{X}^{KL}\right] =\delta^{IL}\mathbf{X}%
^{JK}+\delta^{JK}\mathbf{X}^{IL}-\delta^{IK}\mathbf{X}^{JL}-\delta ^{JL}%
\mathbf{X}^{IK}\text{ },  \label{N10}
\end{equation}%
\begin{equation}
\lbrack\mathbf{X}^{IJ},\mathbf{Y}^{A}]=-\frac{1}{2}\Gamma_{AB}^{IJ}\mathbf{Y}%
^{B}\text{ },  \label{N11}
\end{equation}%
\begin{equation}
\left[ \mathbf{Y}^{A},\mathbf{Y}^{B}\right] =\frac{1}{4}\Gamma_{AB}^{IJ}%
\mathbf{X}^{IJ}\text{ },  \label{N12}
\end{equation}
where the constant matrices $\Gamma^{IJ}$ (with entries $\Gamma_{AB}^{IJ}$)
\ are the generators of transformations of the internal Weyl spinors of
"positive" chirality under the $SO(16)$ rotations (see Appendix). The Lie
algebra values of the fields of interest are\footnote{%
The metric $\mathbf{J}$ is symmetric matrix. Then we restrict the matrices%
\textbf{\ }$\mathbf{J\Omega }_{1},\mathbf{J\Omega}_{2}$ and $\mathbf{JH}$
also to be symmetric. Therefore, from (\ref{N1})-(\ref{N2}) and (\ref{N6})
follows that $\mathbf{\Phi}$ should be symmetric, that is should be a
superposition of the symmetric generators $\mathbf{Y}^{A}$.}:\ 
\begin{equation}
\mathbf{Q}_{\alpha}=\frac{1}{2}q_{\alpha}^{IJ}(x^{0},x^{1})\mathbf{X}^{IJ}%
\text{ },\text{ }\mathbf{P}_{\alpha}=p_{\alpha}^{A}(x^{0},x^{1})\mathbf{Y}%
^{A}\text{ },\text{ }\mathbf{\Phi=}4e^{-i\pi/4}\chi^{A}(x^{0},x^{1})\mathbf{Y%
}^{A}.  \label{N13}
\end{equation}
The physical degrees of freedom in the fields $q_{\alpha}^{IJ}(x^{0},x^{1})$%
\textbf{\ }and $p_{\alpha}^{A}(x^{0},x^{1})$\textbf{\ }are\textbf{\ }those
produced by $128$ space-time physical scalars $\varphi^{A}(x^{0},x^{1})$
which parametrize the $E_{8\left( +8\right) }$ group values of the
orthonormal frame $\mathbf{V=\exp}\left( a^{IJ}\mathbf{X}^{IJ}+\varphi^{A}%
\mathbf{Y}^{A}\right) $. Here the $120$ components $a^{IJ}(x^{0},x^{1})$ are
pure gauge objects which can be chosen in any desirable form by an
appropriate orthogonal rotation of the frame $\mathbf{V}$ (for example $%
a^{IJ}$ can be eliminated, in which case the frame becomes a symmetric
matrix). Then all physical degrees of freedom in the frame current (\ref{N3}%
) come from the scalar fields $\varphi^{A}(x^{0},x^{1})$. The each odd
coefficient $\chi^{A}(x^{0},x^{1})$ (for each fixed value of the index $A$)
in $\mathbf{\Phi}$ represents the two-component Dirac spinor in the
two-dimensional space-time $x^{0},x^{1}$. In this way the numbers of the
physical degrees of freedom (that is the numbers of the arbitrary initial
data) for scalars and spinors are the same ($256$ for scalars since
equations of motion for $\varphi^{A}$ are of the second order in time and $%
256$ for spinors $\chi^{A}$ because they satisfy Dirac-like equations of the
first order in time but have twice more components).

Substituting decomposition (\ref{N13}) into equations (\ref{N7})-(\ref{N8})
and taking into account the relations:%
\begin{equation}
\chi^{A}=\left( 
\begin{array}{c}
\chi_{1}^{A} \\ 
\chi_{2}^{A}%
\end{array}
\right) ,\text{ }\bar{\chi}^{A}=i\left( \chi_{2}^{A},-\chi_{1}^{A}\right) ,%
\text{ }\mathbf{\bar{\Phi}=}4e^{-i\pi/4}\bar{\chi}^{A}\mathbf{Y}^{A},\text{ }%
\bar{\chi}^{A}\gamma^{\alpha}\chi^{B}=-\bar{\chi}^{B}\gamma^{\alpha}\chi ^{A}
\label{N14}
\end{equation}
together with defining commutations (\ref{N10})-(\ref{N12}), we obtain 
\textit{exactly the equations} (\ref{New1})-(\ref{New2}). These equations
are the basic physical equations of motion for the two-dimensional
Nishino-Rajpoot theory. All other fields in this theory (conformal factor $%
\lambda^{2}$ and gravitino) can easily be found in terms of the solutions of
the system (\ref{New1})-(\ref{New2}). To find solutions of these principal
equations one should return to the system (\ref{M21})-(\ref{M24}) and
construct some (as simple as possible) background solutions for the fields $%
\mathbf{J}^{\left( 0\right) },\mathbf{\Omega}_{1}^{\left( 0\right) },\mathbf{%
\Omega}_{2}^{\left( 0\right) },\mathbf{H}^{\left( 0\right) }$ which
parametrize the $E_{8\left( 8\right) }$ group. Then from (\ref{M19}) follows
the background matrix $\mathbf{\Psi}^{\left( 0\right) }\left( \zeta,\eta
,\theta^{1},\theta^{2}\right) $\textbf{, }using which\textbf{\ }one can
integrate the Lax pair (\ref{M15}) to get the corresponding seed solution
for the spectral matrix $\mathbf{\hat{\Psi}}^{\left( 0\right) }\left(
\zeta,\eta,\theta^{1},\theta^{2},s\right) $. All subsequent steps are
closely analogous to the procedure described at the end of the previous
section. The only difference is that now some part of the calculations
should be carried out following the rules of algebra of anticommuting
numbers. After adding to the background a number of solitons we will obtain
the final solutions for the $n$-solitonic $E_{8\left( 8\right) }$ fields $%
\mathbf{J(}\zeta ,\eta),\mathbf{\Omega}_{1}(\zeta,\eta),\mathbf{\Omega}%
_{2}(\zeta,\eta)$ and we can choose any orthonormal frame $\mathbf{V(}%
\zeta,\eta)$ satisfying the relation (\ref{N1}). This frame will give the
matrices $\mathbf{Q}_{\alpha}$ and $\mathbf{P}_{\alpha\text{ \ }}$as is
prescribed by the relation (\ref{N3}) and from (\ref{N13}) we will obtain
the coefficients $q_{\alpha}^{IJ}(x^{0},x^{1})$\textbf{\ }and $%
p_{\alpha}^{A}(x^{0},x^{1}),$ namely those appearing in the equations (\ref%
{New1})-(\ref{New2}). Also with the help of this frame matrix we will
extract from $\mathbf{\Omega}_{1}$ and $\mathbf{\Omega}_{2}$ matrices $%
\mathbf{\Lambda}_{1}$\textbf{\ }and\textbf{\ }$\mathbf{\Lambda}_{2}$ in
accordance with formulas (\ref{N2}) and these matrices will give us the
matrix spinor $\mathbf{\Phi}$ (\ref{N6}). Then the $128$ spinors of interest 
$\chi^{A}$ will follow from the third relation (\ref{N13}). The quantities $%
q_{\alpha}^{IJ}(x^{0},x^{1})\mathbf{,}$ $p_{\alpha}^{A}(x^{0},x^{1}),%
\chi^{A}(x^{0},x^{1})$ calculated in this way are exactly those which will
satisfy automatically equations (\ref{New1})-(\ref{New2}).

Finally it is necessary to stress that one should not try to find a local
supersymmetry directly in the foregoing integrable ansatz for the fields $%
\varphi^{A}$ and $\chi^{A}$ because it corresponds to completely fixed
supersymmetry gauges. It follows from the work \cite{N1} and rules a), b),
c) that after we chose the gauges corresponding to the conformal flat metric 
$g_{\alpha\beta}\left( x^{0},x^{1}\right) $ and to the special form $%
\psi_{\alpha}=\gamma_{\alpha}\psi$ for the gravitino some residual
supersymmetry still remains in the system. This residual freedom can be used
to eliminate the superpartner to the function $\alpha(x^{0},x^{1})$ together
with superpartner to the spectral parameter $s(x^{0},x^{1})$ in the Lax
pair. That's way we can use (without loss of generality) in the super Lax
representation (\ref{M15}) the quantities $\alpha$ and $s$ as the usual even
functions.

In the two-dimensional supergravities such special gauges give rise to the
same miracle as in pure gravity, that is complete separation of the
equations of motion for the matter from gravitational potentials and
gravitinos. Namely due to this fact the models considered in \cite{N1} and
here are integrable. To have the general form for these supergravities one
should apply the backward supersymmetric transformations which
transformations can be extracted from papers \cite{NW} and \cite{NS}.

\section{Appendix}

The sixteen $256\times256$ gamma-matrices $\mathbf{\Gamma}^{I}$ of $SO(16)$
can be chosen \ symmetric and block off-diagonal:%
\begin{equation}
\mathbf{\Gamma}^{I}=\left( 
\begin{array}{cc}
0 & \Gamma_{A\dot{A}}^{I} \\ 
\Gamma_{\dot{A}A}^{I} & 0%
\end{array}
\right) ,\text{ }  \label{A1}
\end{equation}
where $\Gamma_{\dot{A}A}^{I}$ is transposed to $\Gamma_{A\dot{A}}^{I}$. The
Clifford relation $\mathbf{\Gamma}^{I}\mathbf{\Gamma}^{J}+\mathbf{\Gamma}^{J}%
\mathbf{\Gamma}^{I}=2\delta^{IJ}\mathbf{I}$ takes the form:%
\begin{equation}
\sum_{\dot{A}}(\Gamma_{A\dot{A}}^{I}\Gamma_{B\dot{A}}^{J}+\Gamma_{A\dot{A}%
}^{J}\Gamma_{B\dot{A}}^{I})=2\delta^{IJ}\delta_{AB}\text{ },  \label{A2}
\end{equation}%
\begin{equation}
\sum_{A}(\Gamma_{A\dot{A}}^{I}\Gamma_{A\dot{B}}^{J}+\Gamma_{A\dot{A}%
}^{J}\Gamma_{A\dot{B}}^{I})=2\delta^{IJ}\delta_{\dot{A}\dot{B}}\text{ }.
\label{A3}
\end{equation}

It is known that the real solutions of these equations exist and can be
constructed, for example, in the way analogous to what has been done in
appendix A of the paper \cite{GS} for the group $SO(8)$.

The $SO(16)$ generators of spinorial transformation $\mathbf{\Gamma}^{IJ}=%
\frac{1}{2}(\mathbf{\Gamma}^{I}\mathbf{\Gamma}^{J}-\mathbf{\Gamma}^{J}%
\mathbf{\Gamma}^{I})$ are block diagonal: 
\begin{equation}
\mathbf{\Gamma}^{IJ}=\left( 
\begin{array}{cc}
\Gamma_{AB}^{IJ} & 0 \\ 
0 & \Gamma_{\dot{A}\dot{B}}^{IJ}%
\end{array}
\right) ,  \label{A4}
\end{equation}
where%
\begin{equation}
\Gamma_{AB}^{IJ}=\frac{1}{2}\sum_{\dot{A}}\left( \Gamma_{A\dot{A}%
}^{I}\Gamma_{B\dot{A}}^{J}-\Gamma_{A\dot{A}}^{J}\Gamma_{B\dot{A}}^{I}\right)
,  \label{A5}
\end{equation}%
\begin{equation}
\Gamma_{\dot{A}\dot{B}}^{IJ}=\frac{1}{2}\sum_{A}\left( \Gamma_{A\dot{A}%
}^{I}\Gamma_{A\dot{B}}^{J}-\Gamma_{A\dot{A}}^{J}\Gamma_{A\dot{B}}^{I}\right)
.  \label{A6}
\end{equation}
The blocks $\Gamma_{AB}^{IJ}$ and $\Gamma_{\dot{A}\dot{B}}^{IJ}$ generates
the transformations of the internal Weyl spinors of "positive" and
"negative" chiralities respectively. Both components $\Gamma_{AB}^{IJ}$ and $%
\Gamma _{\dot{A}\dot{B}}^{IJ}$ are antisymmetric under interchange of the
upper indices $I,J$ as well as lower indices $A,B$ and $\dot{A},\dot{B}$.

\end{document}